\documentclass[namedreferences]{kluwer}
\usepackage{graphicx}

\def\flat{type\,V}

\def\ser{S\'ersic }
\begin{document}
\begin{article}
\begin{opening}
\title{New insights into the photometric structure of Blue Compact
Dwarf Galaxies from a deep Near--Infrared study}

\author{K.G. \surname{Noeske}\email{knoeske@uni-sw.gwdg.de}}
\author{P. \surname{Papaderos}} 
\author{L.M. \surname{Cair\'os}} 
\author{K.J. \surname{Fricke}} 
\institute{Universit\"ats-Sternwarte G\"ottingen, Germany
}                               




\runningtitle{NIR observations of BCDs}
\runningauthor{K. Noeske et al.}



\begin{abstract} 
We present deep Near--Infrared (NIR) imaging of Blue Compact Dwarf
Galaxies (BCDs), allowing for the first time to derive and systematize
the NIR structural properties of their stellar low--surface brightness
(LSB) host galaxies. Compared to optical data, NIR images, being less
contamined by the extended stellar and ionized gas emission from the
starburst, permit to study the LSB host galaxy closer to its center.
We find that radial surface brightness profiles (SBPs) of the LSB
hosts show at large radii a mostly exponential intensity distribution,
in agreement with previous optical studies. At small to intermediate
radii, however, the NIR data reveal an inwards flattening with respect
to the outer exponential slope (``type V SBPs'', Binggeli \& Cameron
1991) in the LSB component of more than one half of the sample BCDs.
This result may constitute an important observational constraint to
the dynamics and evolution of BCDs.  We apply a modified exponential
fitting function (Papaderos et al. 1996a) to parametrize and
systematically study type V profiles in BCDs. A S\'ersic law is found
to be less suitable for studying the LSB component of BCDs, since it
yields very uncertain solutions.
\end{abstract}




\end{opening}

\section{Introduction:}

Studies of Blue Compact Dwarf Galaxies (BCDs) are crucial for
understanding the starburst--driven evolution of gas--rich low--mass
extragalactic systems at low and high redshift. Particularly important
for the study of BCDs is the information on their stellar low-surface
brightness (LSB) host galaxy (Loose \& Thuan \citeyear{loose86}).
This evolved component, present in most types of local dwarf galaxies,
contains the bulk of the stellar mass of a BCD and is most probably
dynamically relevant. It is therefore likely that its structure (e.g.,
mass density distribution) influences the global star formation
process in a BCD.
However, the intensity distribution of the LSB component is uncertain
close to its center, where it is most required to understand the
centrally concentrated star--forming (SF) activity of typical BCDs.
Previous optical studies of these systems have been restricted to the
LSB periphery ($\gsim$2 exponential scale lengths) where the emission
of the starburst becomes negligible.
At NIR wavelengths, the fractional flux contribution of the starburst
is comparatively low, allowing to study the LSB hosts of BCDs closer
to their center.
We present results from the first large BCD sample, comprising $>$ 30
objects of all mor\-pho\-lo\-gi\-cal sub\-clas\-ses, which was
observed sufficiently deep in the NIR to sys\-te\-ma\-ti\-cal\-ly
study the structural properties of the LSB hosts. The complete
analysis of the sample is presented in Noeske et
al. (\citeyear{noeske02}) and Cair\'os et al. (\citeyear{cairos02}).
%
\section{Observations, data reduction \& analysis}

NIR images were observed at the Calar Alto 3.6m telescope with OME\-GA
PRIME, at the ESO NTT with SOFI and at the ING 4.2m WHT with the
INGRID camera. Typical on-object integration times are $\gsim$ 20 min
in $J$, $\gsim$ 25 min in $H$ and $\gsim$ 30 min in $K$. All
instruments were equipped with 1k$\times$1k pixel NIR arrays, which improve
both the observing efficiency, and the quality of the background
correction. The latter is essential for surface photometry studies,
and could be further refined through a data reduction software package
we developed, which complements standard NIR reduction techniques by
new correction procedures.
Surface brightness and color profiles (Fig. 1), derived as described
in Papaderos et al. (2002), reach limiting surface brightnesses of
23.5 - 25.5 mag/$\Box ''$ in $J$ and 22 -- 24 mag/$\Box ''$ in $H$
and $K$.

\begin{figure*}
\centerline{%
\hspace*{0.0mm}\includegraphics[width=5.5cm,clip=]{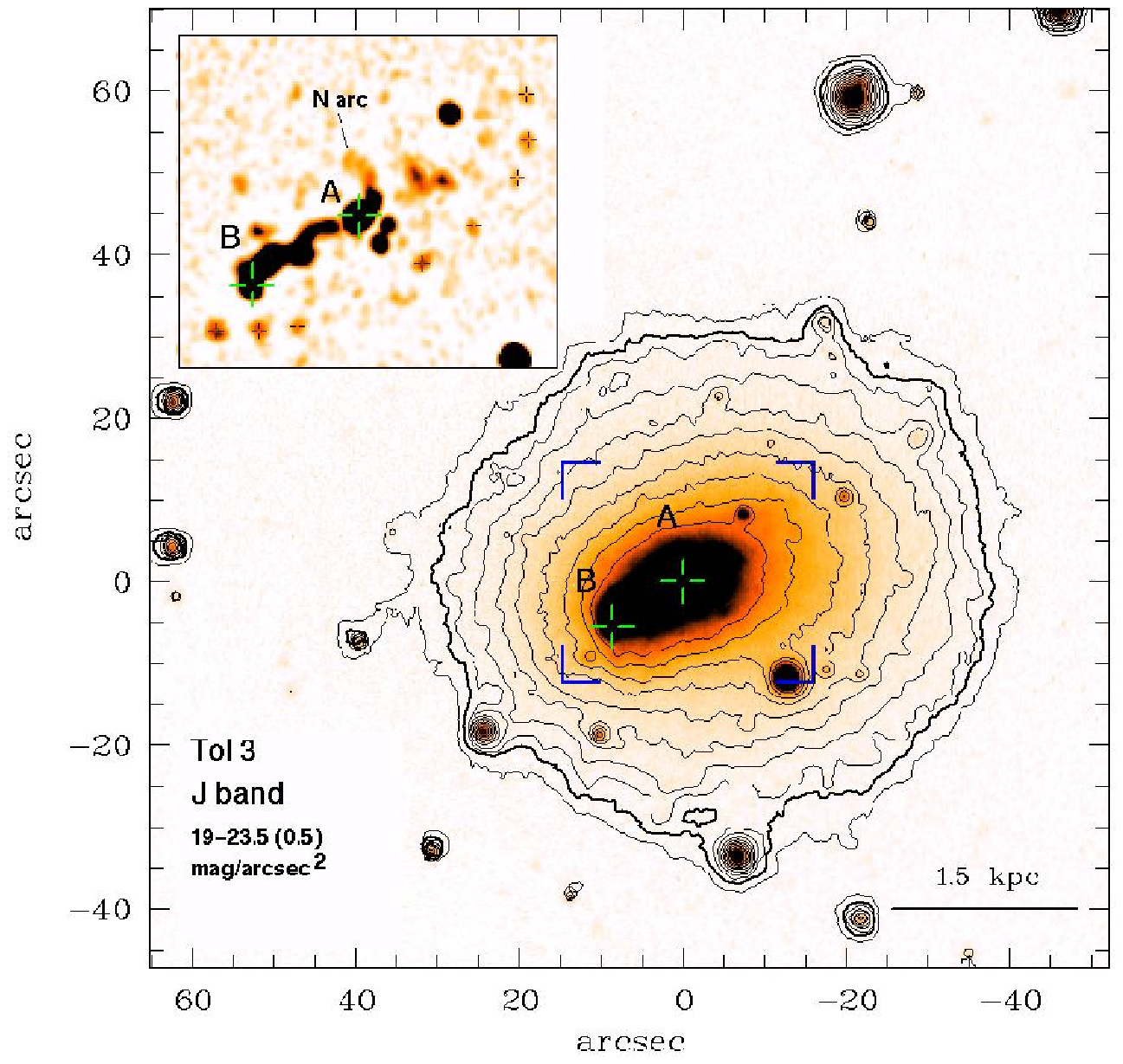}
\hspace*{5.0mm}\includegraphics[width=5.5cm,clip=]{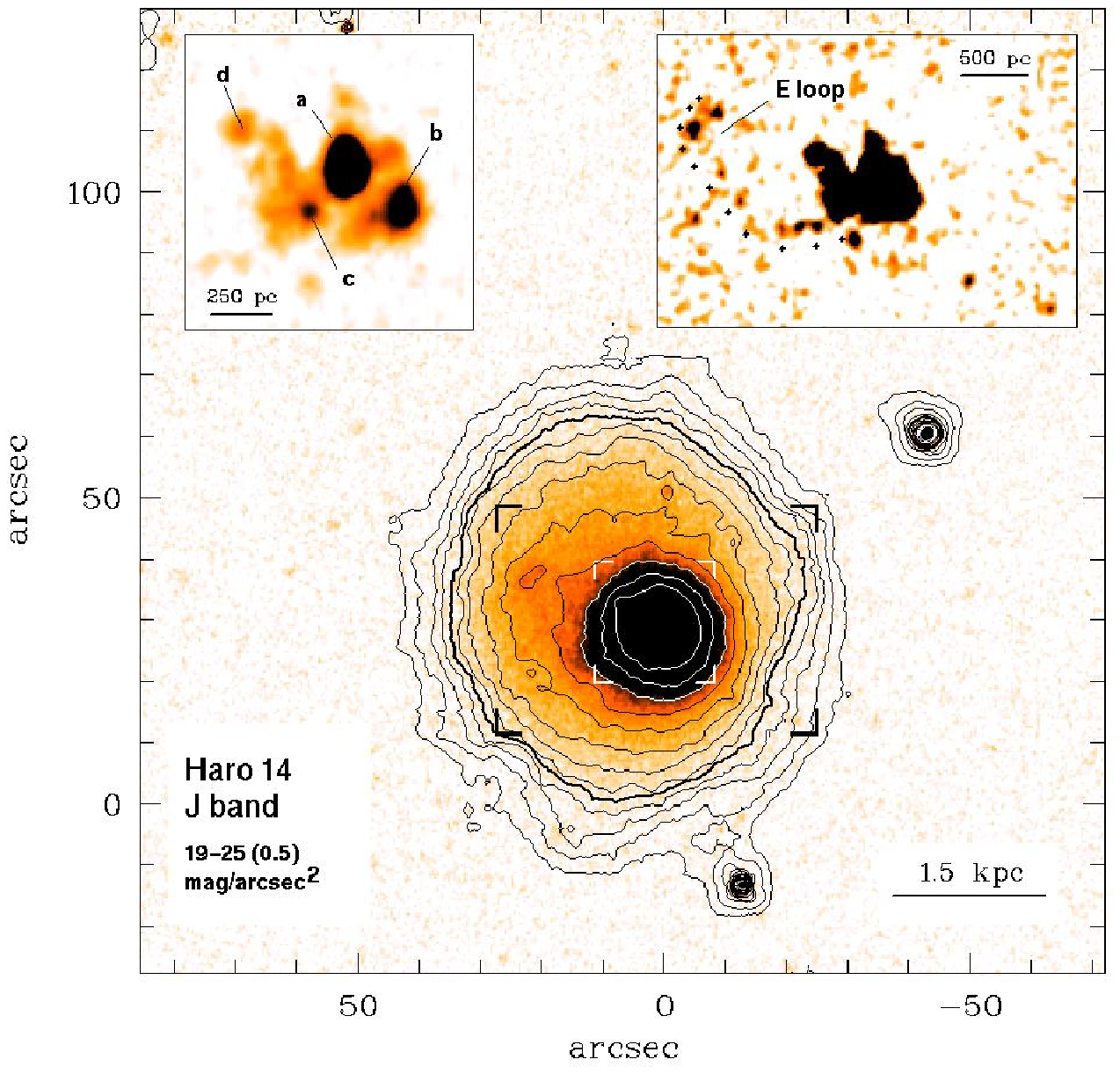}
}
\centerline{%
\hspace*{1.5mm}\includegraphics[angle=270,width=5.3cm,clip=]{noeske.fig3.eps}
\hspace*{6.0mm}\includegraphics[angle=270,width=5.3cm,clip=]{noeske.fig4.eps}
}
\centerline{%
\hspace*{2.75mm}\includegraphics[angle=270,width=5.3cm,clip=]{noeske.fig5.eps}
\hspace*{5.95mm}\includegraphics[angle=270,width=5.35cm,clip=]{noeske.fig6.eps}
}
\caption{{\bf Top:} $J$ band images and $J$ isophotes at intensity
levels as indicated in the legends. The insets show contrast--enhanced
blow--ups of the regions indicated by the brackets. {\bf center:}
Surface brightness profiles in $J$, $H$ and $K$. Heavy grey lines show
fits to the LSB component by exponential (Tol 3) and modified
exponential (Haro 14) functions. {\bf bottom:} Color profiles. Thick
lines to the right indicate the mean colors of the LSB host galaxy.}
\end{figure*}
\section{Results}

{\bf (i)} Contrast-enhanced NIR maps uncover numerous morphological
details within the SF component (Fig. 1), which may hold
clues to the history and spatial progression of SF
activities in BCDs. In some cases, a combination of NIR and optical
data indicates appreciable and non-uniform dust absorption on a
spatial scale as large as $\sim$1 kpc.\\[0.3em]
{\bf (ii)} Surface brightness profiles (SBPs) of our sample BCDs can
at large galactocentric radii $R^{\star}$ be well approximated by an
exponential fitting law (Fig. 1), in agreement with previous evidence
from optical surface photometry. Also the exponential scale lengths $\alpha$
derived in the optical and NIR are in mutual agreement, implying minor
optical--NIR color gradients within the LSB component.  For the majority
of the LSB host galaxies in our sample we derive optical-NIR colors
indicative of an evolved stellar population of subsolar metallicity.\\[0.3em]
{\bf (iii)} On a galactocentric radius of 1\dots 3$\alpha$, the underlying 
LSB component of several BCDs shows a
conspicuous intensity depression with respect to the purely
exponential slope derived for larger radii (Fig. 1, right). Such
inwards flattening exponential SBPs, classified ``\flat '' in Binggeli
\& Cameron (\citeyear{binggeli91}), are not uncommon in other low-mass
systems, e.g. dIs. Our NIR data reveal signatures of a \flat\ SBP in
the LSB host of more than one half of the sample BCDs, suggesting a
higher frequency of such profiles than previously indicated by optical
studies.  This discrepancy is likely attributable to the extended
starburst emission in optical wavelengths, which can readily mask a
flattening in the inner part of the underlying stellar LSB
component. A possible high frequency of \flat\ profiles among dwarf
galaxies is not expected to significantly change a structural
dichotomy of the LSB component of BCDs from other types of dwarf
galaxies (Papaderos et al. \citeyear{papaderos96b}).  However, it
could have important implications for our view about BCDs, as it would
significantly increase the estimated starburst-to-LSB luminosity
fraction, and therefore the amount of photometric fading of these
systems, once the starburst activity has terminated. This information
is crucial for, e.g., establishing or discarding the hypothesis of
faint dwarf spheroidals being the evolutionary endpoints of BCDs.  In
the same way, a \flat\ intensity distribution would impose new
observational constraints to the total stellar mass
and its intrinsic luminosity density profile within the LSB component
of BCDs.\\[0.3em]
{\bf (iv)} The physical origin of \flat\ SBPs in dwarf galaxies is to
date not understood. We find that such SBPs can be well approximated
by a modified exponential fitting formula (Papaderos et
al. \citeyear{papaderos96a}). This empirical model is found to be
suitable to systematize the structural properties of BCDs, and for
a meaningful decomposition of the SBPs of these systems into the LSB
and the starburst component.  A S\'ersic law can also yield good fits
to \flat\ profiles, albeit small systematic residuals.  However, the
practical applicability of the \ser  law to the LSB emission of BCDs
is limited by the strong non-linear coupling of its free parameters,
and the extreme sensitivity of the achieved solutions to, e.g., small
uncertainties in the sky subtraction and SBP derivation.
%
\begin{acknowledgements}
KGN acknowledges support from the Deutsche Forschungsgemeinschaft
(DFG) Grants FR325/50-1 and FR325/50-2. LMC has received support from
the EC Grant HPMF-CT-2000-00774.
\end{acknowledgements}

\end{article}
\end{document}